# Uni-Electrolyte: An Artificial Intelligence Platform for Designing Electrolyte Molecules for Rechargeable Batteries

Xiang Chen[a,b,*], Mingkang Liu[b], Shiqiu Yin[b], Yu-Chen Gao[a], Nan Yao[a], and Qiang Zhang[a,*]

[a]Tsinghua Center for Green Chemical Engineering Electrification (CCEE), Beijing Key Laboratory of Green Chemical Reaction Engineering and Technology, Department of Chemical Engineering, Tsinghua University, China

[b]AI for Science Institute, Beijing 100080, China

X.C. and M.L. contribute equally to this work.

**Corresponding authors.** E-mails: xiangchen@mail.tsinghua.edu.cn;
zhang-qiang@mails.tsinghua.edu.cn

**Abstract:**

Electrolyte is a very important part of rechargeable batteries such as lithium batteries. However, the electrolyte innovation is facing grand challenges due to the complicated solution chemistry and infinite molecular space (>$10^{60}$ for small molecules). This work reported an artificial intelligence (AI) platform, namely Uni-Electrolyte, for designing advanced electrolyte molecules, which mainly includes three parts, i.e. EMolCurator, EMolForger, and EMolNetKnittor. New molecules can be designed by combining high-throughput screening and generative AI models from more than 100 million alternative molecules in the EMolCurator module. The molecule properties including frontier molecular orbital information, formation energy, binding energy with a Li ion, viscosity, and dielectric constant, can be adopted as the screening parameters. The EMolForger, and EMolNetKnittor module can predict the retrosynthesis pathway and reaction pathway with electrodes for a given molecule, respectively. With the assist of advanced AI methods, the Uni-Electrolyte is strongly supposed to discover new electrolyte molecules and chemical principles, promoting the practical application of next-generation rechargeable batteries.

# 1. Introduction

Rechargeable batteries especially lithium (Li) ion batteries (LIBs) have been widely applied in modern society, from electronic devices, electric vehicles, and smart grids to low-altitude industry. These wide applications increasingly put forward demanding requirements for next-generation batteries with high safety, high energy density, high power density, long lifespan, or wide temperature window (3H1L1W). Among various battery technology innovations, designing advanced electrolyte molecules has been strongly considered as one of the most promising approach due to the significant role of electrolytes in stabilizing battery interfaces and regulating battery performance [1-5]. Besides, new electrolytes can be directly adopted in current battery manufacturing without a huge change of the equipment.

Looking backing to the history of LIB developments, the use of ethylene carbonate (EC)-based electrolyte is definitely a milestone due its irreplaceable role in forming a stable solid electrolyte interphase (SEI) on graphite anode [6-8]. EC-based ester electrolytes still dominate the current LIB battery market after more than thirty-year developments. However, ester molecules are very active towards Li metal anodes, which possesses the largest specific capacity as well as a very low electrode potential and is most promising choice for constructing ultrahigh energy density batteries [9]. Beyond ester molecules, ether solvents such as 1,2- dimethoxyethane (DME) has widely applied to Li metal batteries due to its relatively high stability against Li metal anode and good Li salt solubility. High-concentration and localized high-concentration electrolytes have wide further constructed mainly based on DME solvents and have demonstrated promising electrochemical performances [5]. Besides, many electrolyte additives have been proposed to improve battery performance. For example, vinyl carbonate (VC) and fluoroethylene carbonate (FEC) can be adopted to produce a polymeric layer on electrode surface. Trimethyl phosphate (TMP) and biphenyl can be adopted as flame retardants and overcharge protective additive, respectively. All above examples ensure the important role of new electrolyte innovation in battery developments [10-11].

Tremendous electrolyte molecules have been explored for rechargeable batteries mainly through a conventional trial-and-error approach while only dozens of them are applied to current commercialized batteries. The time-consuming and low-efficiency approach for search advanced electrolytes is a major challenge for promoting next-generation batteries during the following years. Recently, the rising of artificial intelligence (AI) technology afford new and promising chance for the frontier research of chemistry and materials science [12-17]. Especially, the Nobel prices in both physics and chemistry 2024 are rewarded to AI-related filed due to the great success of AI. In the electrolyte filed, previous studies particularly focused on developing machine learning potentials to improve the simulation time and length scales while maintain a first-principles accuracy [18-20]. Chen and authors developed data–knowledge-dual-driven and explainable machine learning models to directly predict electrolyte properties and establish molecular structure–function relationship, respectively. Beside, several electrolyte molecules have been predicted by combining machine learning models and high-throughput screening [21-22]. Although great successes have been demonstrated for applying AI to probe electrolyte chemistry and design promising electrolyte molecules, a platform that integrates advanced AI and domain knowledge for designing advanced battery electrolytes has not been reported.

In this contribution, we reported the first AI platform, namely Uni-Electrolyte, for designing electrolyte molecules for rechargeable batteries. The Uni-Electrolyte integrates advanced AI algorithms for electrolyte molecule design, mainly including three modules. The EMolCurator module can assist design new molecules through high-throughput screening from an embedded electrolyte database or an upload molecular database, and designing new molecules beyond previous database using AI-based generative models. When target molecule is ensured, the EMolForger module can predict the retrosynthesis pathways and reaction conditions, which are helpful to synthesize the predicted molecules. Last but not least, the EMolNetKnittor module can unveil the reaction pathway of the forming the SEI production, which is widely supposed as a critical factor in stabilizing the electrolyte–electrode interphase. Collectively, the Uni-Electrolyte is supposed to discover new electrolyte molecules including solvents and additives and further promote the practical application of next-generation batteries.

## 2. Methods

*2.1 The Framework of Uni-Electrolyte*

The Uni-Electrolyte aims to expedite the discovery, synthesis, and analysis of novel electrolyte materials using advanced AI technology (Figure 1). The platform mainly includes three interconnected modules: EMolCurator, EMolForger, and EMolNetKnittor. The functions of each module will be introduced first and then their frameworks.

**EMolCurator** is the foundational module, working for the rational design of potential electrolyte molecules. It leverages a large database constructed by both density functional theory (DFT) calculations and molecular dynamics (MD) simulations to train and deploy advanced machine learning models, including quantitative structure–property relationship (QSPR) and AI-based generative models. These models enable the accurate prediction of critical properties such as the highest occupied molecular orbital (HOMO), the lowest unoccupied molecular orbital (LUMO), binding energy with a Li ion, dielectric constant ($\varepsilon$), and viscosity ($\eta$). Furthermore, user-friendly interfaces are provided to facilitate efficient database screening and querying to rapidly identify promising candidates.

**EMolForger** module aims to predict the synthesis pathway and corresponding reaction conditions of the as-predicted molecule candidates that can not be brought directly from markets. It incorporates a sophisticated synthetic route planner and an AI-powered single-step retrosynthesis predictor, both trained on a comprehensive electrolyte database. The module can assist in identifying efficient and cost-effective synthesis pathways, significantly accelerating the translation of computational designs into laboratory realities.

**EMolNetKnittor** aims to unveil the working mechanism of as-design electrolyte molecules in batteries, focusing on the formation mechanism of the SEI. It employs advanced computational techniques to perform SEI-product analysis, providing valuable insights into the electrochemical performance of the designed electrolytes.

This module aids in understanding the complex interfacial processes that occur at the electrode-electrolyte interface, enabling the optimization of electrolyte formulations.

By integrating the three modules, Uni-Electrolyte offers a powerful and versatile platform for accelerating the discovery and development of advanced electrolyte materials. This innovative approach has the potential to revolutionize the field of battery research, leading to the development of next-generation batteries with superior energy density, power density, and cycle life.

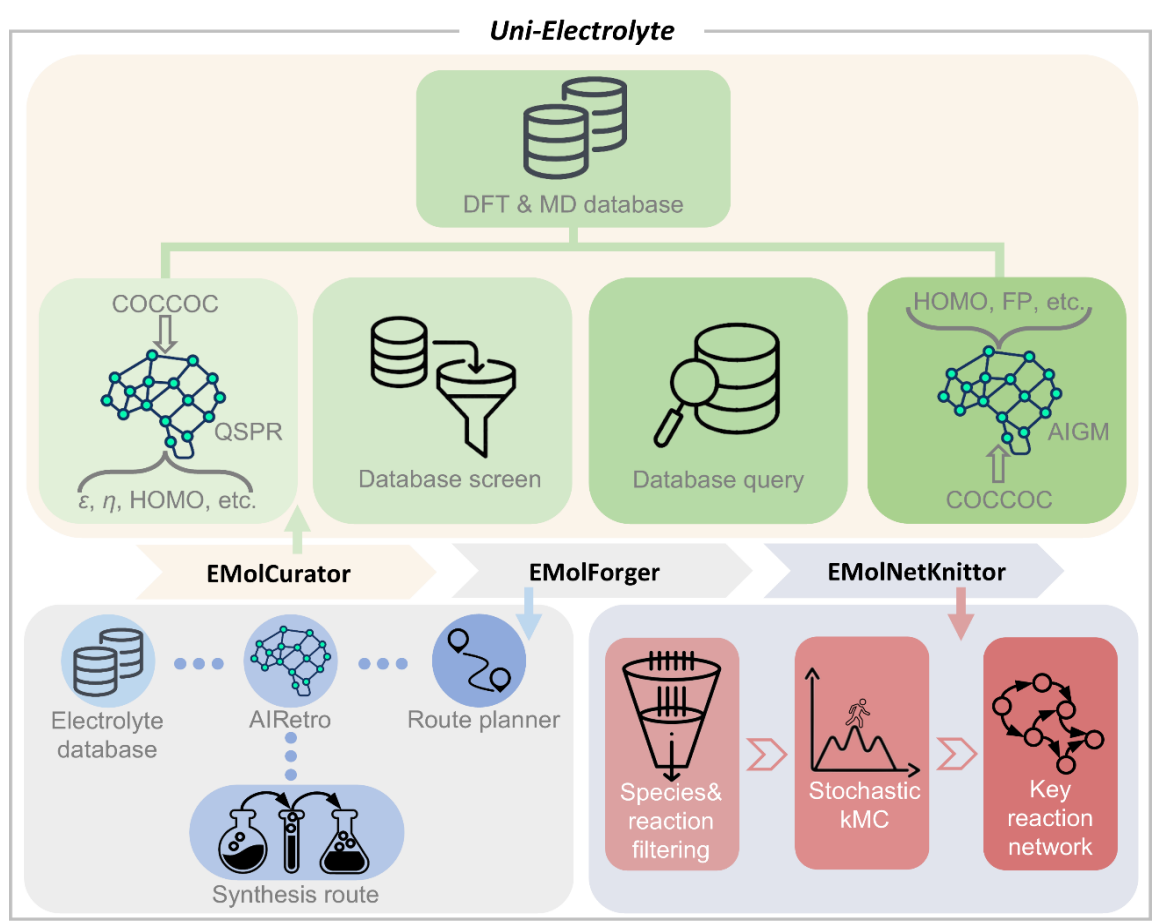

**Figure 1.** Schematic representation of the Uni-Electrolyte platform with three modules. The 'EMolCurator' module aims to design new electrolyte molecules. Based embedded electrolyte database, QSPR and AI-based generative models were trained. The 'EMolForger' module can predict the synthesis pathways and corresponding reaction conditions of potential electrolyte molecules. It was built with a synthetic route planner and AI-based single-step retrosynthesis predictor. The 'EMolNetKnittor' module

assesses the filtered electrolyte species and reaction database to propose chemical reaction networks and perform SEI-product analysis.

*2.2 EMolCurator: AI-Assisted Molecular Design Framework*

To address the limitations of traditional molecular design approaches, the EMolCurator enables efficient and precise molecular design. EMolCurator leverages a combination of advanced AI techniques and quantum mechanics calculations to explore vast chemical spaces and identify promising molecular candidates.

*2.2.1 Molecular Property Prediction*

The cornerstone of EMolCurator is a robust QSPR model, trained on a comprehensive electrolyte dataset constructed from DFT and MD calculations. The model accurately predicts various molecular properties, including binding energy, HOMO energy, LUMO energy, viscosity, dielectric constant, and other relevant properties, from both 2D and 3D molecular representations (Figure 2a).

*2.2.2 Multi-criteria Screening*

A multi-criteria screening process is employed on a pre-designed database of electrolyte molecules (Figure 2b). The process involves filtering molecules based on multiple property criteria, such as desired HOMO-LUMO gap, binding energy, or specific structural features. By combining these criteria, a subset of molecules could be identified that are most likely to meet the desired specifications, accelerating the design process.

*2.2.3 Molecular Similarity Search*

A similarity search algorithm that leverages QSPR-predicted molecular properties was built upon a vector database. A pgvector-enhanced PostgreSQL backend was constructed to store molecules and their corresponding property vectors. These vectors, representing a combination of physical and electrical properties, enable efficient similarity searches. When a query molecule is input, it is either retrieved from the database or its properties are predicted on-the-fly using a state-of-the-art model. The query vector is then compared to database vectors using a similarity metric, such as cosine similarity, to identify molecules with similar property profiles. This approach expands the search space while maintaining focus on relevant chemical regions.

*2.2.4 Molecular Generation*

To overcome the limitations of relying solely on existing datasets, an AI-driven molecular generation module was built. This module, trained on the same DFT and MD dataset, can generate novel molecules with targeted properties. Three specific generation tasks are considered as follows.

(1) HOMO–LUMO gap optimization: Generating molecules with a specific HOMO–LUMO gap to achieve desired electronic properties.

(2) Binding energy minimization: Generating molecules with low binding energy to specific target molecules.

(3) Structural Design: Generating molecules with similar structures to user-provided templates.

To ensure the practicality of the generated molecules, a rigorous filtering pipeline is implemented. For example, the stability of molecules, which can be indicated by the formation energy; the similarity of new molecules compared with existing molecules in the dataset; and the synthesizability of new molecules.

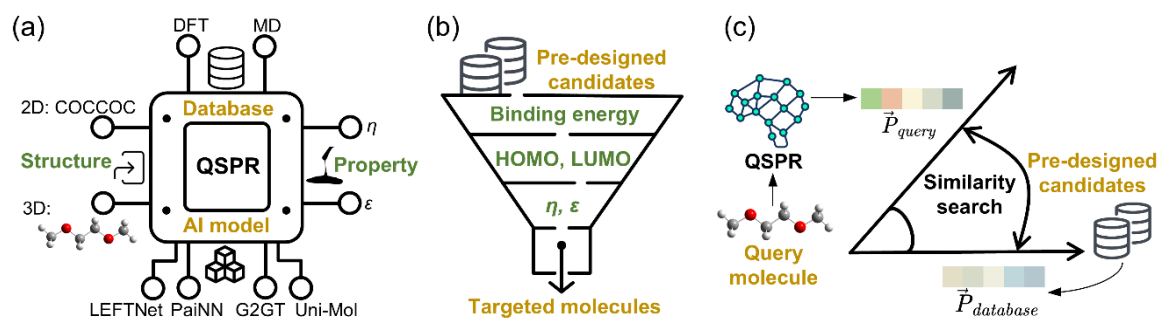

**Figure 2.** Three functions of the EMolCurator module. (a) The QSPR model is benchmarked and trained on the DFT and MD databases. It intakes 2D or 3D molecular graphs and outputs their properties. (b) Pre-designed candidate electrolyte molecules are screened with respect to user-defined intervals. (c) Query similar-properties molecules with the vector database, the queried vector itself is composed of the predicted properties from the QSPR model.

The entire molecular design process is iterative, with user-defined targets guiding the generation and filtering steps (Figure 3). The workflow can be repeated until convergence is achieved, leading to the identification of optimal molecular candidates. By combining the above functions, the EMolCurator module can provide a powerful tool for accelerating the discovery of novel electrolyte molecules with tailored properties. The AI-driven nature of the framework enables rapid exploration of vast chemical spaces, leading to the identification of promising candidates that may be difficult to discover through traditional methods.

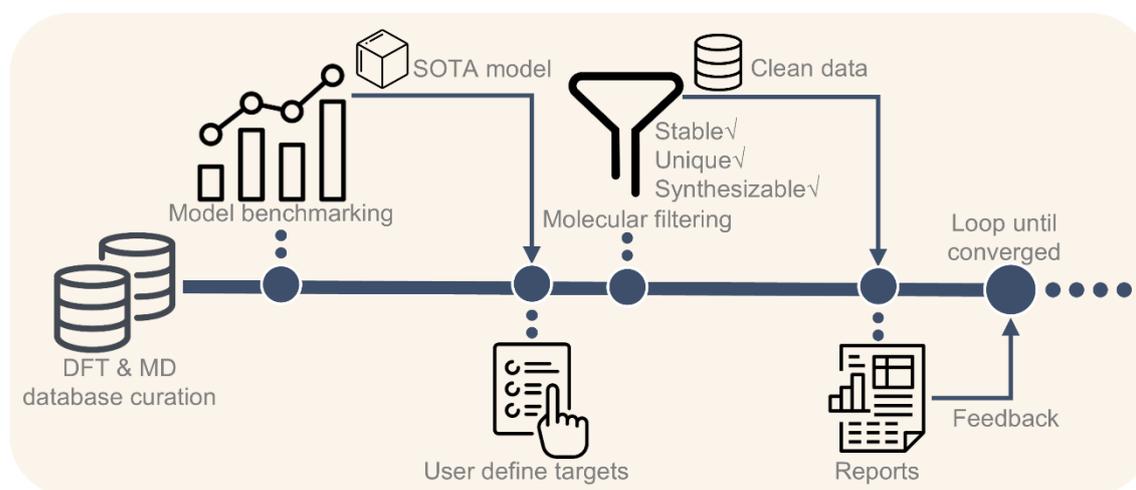

**Figure 3.** The AI-driven molecule generation workflow. Benchmarked on DFT & MD databases, the SOTA model can be guided with user-defined targets. An automated cleaning pipeline is built to identify stable, unique and synthesizable molecules. The entire workflow can be iterated until convergence is reached.

*2.3 EMolForger: AI-Powered Retrosynthetic Analysis Module*

Retrosynthetic analysis is a critical bridge between theoretical design and experimental implementation, which mainly relied on human expertise and intuitive chemistry knowledge previously. However, the increasing complexity of electrolyte molecules, coupled with the vast chemical space to be explored, demands more sophisticated and automated approaches. This section introduces the EMolForger module, an AI-powered retrosynthetic analysis platform specifically designed to address these challenges.

The EMolForger module represents a significant advancement in the application of artificial intelligence to retrosynthetic analysis, with specific optimization for electrolyte molecules. The architecture integrates two sophisticated AI components that function in concert to enhance synthetic route planning (Figure 4). The G2GT One-Step AIRetro Predictor, built upon graph neural networks [23], demonstrates exceptional capability in identifying and evaluating feasible one-step retrosynthetic transformations. Through extensive training on comprehensive chemical reaction databases, the predictor has developed robust algorithms for assessing chemical plausibility, establishing a reliable foundation for synthesis planning. The Askcos Route Planner [24] builds upon these initial predictions to optimize multi-step synthetic pathways. This component conducts detailed analyses of crucial reaction parameters, including solvent selection, catalyst optimization, and temperature conditions, while prioritizing both synthetic efficiency and economic viability.

The practical implementation of EMolForger extends beyond theoretical route planning through its sophisticated reagent analysis capability. The system provides comprehensive specifications for chemical reagents while conducting detailed cost analyses, enabling researchers to make well-informed decisions regarding reagent selection and alternatives. The effectiveness of EMolForger is substantially enhanced by its domain-specific optimization, with models fine-tuned using reaction datasets specifically relevant to electrolyte molecule synthesis. The focused training approach ensures exceptional accuracy within the specialized domain of electrolyte chemistry, addressing the unique challenges and requirements of this field.

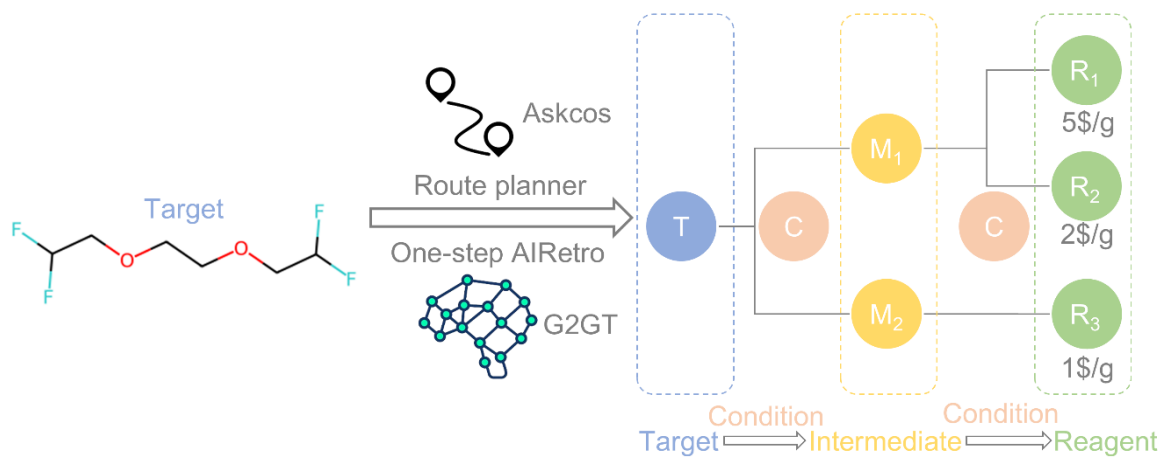

**Figure 4.** The illustration of the retrosynthesis module. The module includes two AI components, *i.e.* the G2GT one-step retrosynthesis predictor and the Askcos synthetic route planner. During inference, the module proposes purchasable starting reagents and potential intermediates, along with detailed reaction conditions and prices.

*2.4 EMolNetKnittor: Comprehensive SEI Formation Analysis Platform*

Following the molecular design strategies and synthetic planning capabilities discussed in previous sections, understanding the behavior and decomposition pathways of successfully synthesized electrolyte molecules becomes crucial. The solid electrolyte interphase (SEI) formation, a critical process affecting battery performance, requires sophisticated analytical tools for comprehensive investigation. Here we present the EMolNetKnittor module, an advanced platform that bridges the gap between molecular synthesis and performance evaluation by enabling detailed analysis of SEI formation mechanisms.

EMolNetKnittor employs a two-pronged approach to analyze SEI formation. Initially, the input electrolyte molecule is queried against a comprehensive built-in database. If relevant species and reactions are identified, the platform utilizes stochastic kinetic Monte Carlo (kMC) simulations to construct a detailed reaction network. This network facilitates the identification of key products and their formation pathways.

However, if the database lacks information about the input molecule, the EMolNetKnittor provides an on-the-fly database-building capability, which involves fragmenting and recombining the input molecule to generate potential species.

Subsequently, one-step reactions are enumerated among these species, and rigorous filters are applied to ensure the validity of both species and reactions. This process results in a tailored database specific to the electrolyte molecule under investigation.

The analytical capabilities of EMolNetKnittor encompass several critical domains. The platform enables the prediction of major decomposition products while elucidating their formation mechanisms and stability parameters. Through comprehensive pathway analysis, the system identifies rate-determining reaction steps and evaluates competing mechanistic routes. Furthermore, the platform quantifies SEI layer composition, formation kinetics, and thermodynamic stability, correlating these parameters with electrochemical performance metrics.

To address the limitations of existing tools like HiPRGen [25], which support only a subset of the LiBE [26] electrolyte database, the database in the EMolNetKnittor module was expanded to encompass the entire LiBE dataset. The expanding dataset enables the analysis of electrolyte molecules containing a wider range of elements, such as F, N, P, and S. Additionally, the database construction process has been automated, empowering users to add custom molecules and further broaden the application scope of EMolNetKnittor.

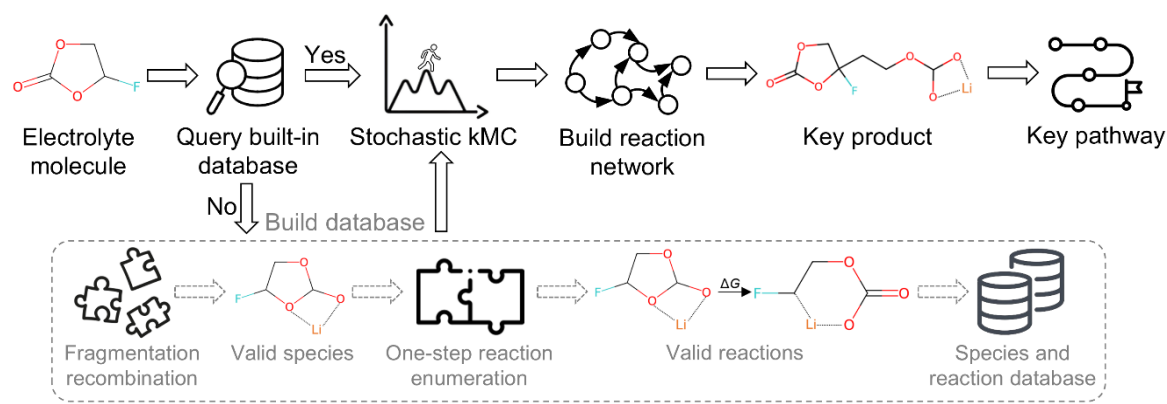

**Figure 5.** Illustration of the SEI-analysis module. The input electrolyte molecule is queried against a built-in database at first. If related species and reactions are found, this module utilizes stochastic kMC simulations to build a reaction network. This network then allows the identification of key products and their formation pathways. However, if no matches are found in the database, the module offers an on-the-fly

database building option, which takes the queried molecule as input and generates species through fragmentation and recombination, followed by reaction enumeration. Note that filters are applied to ensure the validity of both generated species and reactions.

## 3. Results and discussion

*3.1 QSPR Model Performance*

The QSPR models were assessed to establish their capacity to predict critical physical and electronic properties, namely dielectric constant ($\epsilon$), viscosity ($\eta$), HOMO, LUMO, and the binding energy with a Li ion. This initial evaluation sets the stage for a deeper exploration of the interplay between model architecture, dataset characteristics, and property-specific performance.

To ensure the rigor of these benchmarks, the DFT&MD database was partitioned into training, validation, and test datasets in a 3.17:0.37:1 ratio (34,358:3,435:9,056). Further, the test dataset was split into Independent and Identically Distributed (IID) and Out-of-Distribution (OOD) subsets around 1:1 ratio (4,604:4,452), enabling a nuanced evaluation of the models' generalizability. OOD molecules were identified using the Bemis-Murcko scaffold grouping method [27], which classifies molecules based on the rarity of their molecular backbones. This methodology ensured a meaningful comparison of model performance across familiar and novel chemical spaces.

The benchmarking spanned both 2D and 3D QSPR models, revealing distinct patterns in their predictive capabilities. Among the 2D models, those pre-trained on large public datasets showed clear advantages. G2GT [23], pre-trained on the USPTO [28] chemical reaction database, consistently outperformed Uni-Mol [29], which was pre-trained on QM9 [30], by approximately 2%. The results suggest that the chemical reactivity data embedded in USPTO better aligns with the prediction of physical properties like dielectric constant and viscosity.

In contrast, the 3D QSPR models demonstrated the pivotal role of geometric representation in capturing molecular features. Equivariant models such as LEFTNet [31] and PaiNN [32] significantly outperformed the invariant SchNet [33] model, highlighting the importance of symmetry-aware architectures. Notably, powerful invariant models such as SphereNet [34] were not considered here due to their extremely high computational cost as mentioned in previous reports [31]. LEFTNet emerged as the top-performing model, surpassing PaiNN by 4%. This superiority is likely attributable to its architectural enhancements, such as the Local Substructure Encoding (LSE) and Frame Transition Encoding (FTE) modules, which enable comprehensive encoding of both local and global molecular information.

The relative performance of these models is summarized in Table 1, providing a detailed comparison of their Mean Absolute Errors (MAE) on IID and OOD datasets. This analysis reveals not only the strengths of specific models but also underscores the challenges posed by rare molecular scaffolds. These insights naturally lead to the next focus: property-targeted generation.

**Table 1.** Comparison of QSPR performances. $\varepsilon$: Dielectric constant; $\eta$: Viscosity (mPa*s).

| Properties | | Methods | | | | | |
|---|---|---|---|---|---|---|---|
| | | Uni-Mol | G2GT | Graphormer | SchNet (RDKIT) | LEFTNet (RDKIT) | PAINN (RDKIT) |
| IID MAE | $E_b$ (eV) | **0.132** | 0.137 | 0.145 | 0.154 | <u>0.135</u> | 0.142 |
| | $\varepsilon$ | **3.16** | <u>3.18</u> | 3.511 | 3.562 | 3.271 | 3.451 |
| | $\eta$ | 6.22 | <u>6.094</u> | 6.79 | 6.461 | **5.977** | 6.26 |
| | HOMO (eV) | 0.153 | **0.147** | **0.147** | 0.194 | 0.164 | 0.18 |
| | LUMO (eV) | 0.224 | <u>0.205</u> | **0.199** | 0.258 | 0.209 | 0.23 |
| OOD MAE | $E_b$ (eV) | 0.162 | **0.16** | 0.176 | 0.181 | <u>0.161</u> | 0.166 |
| | $\varepsilon$ | 3.32 | 3.31 | 3.665 | 3.427 | **3.271** | <u>3.291</u> |
| | $\eta$ | 13.61 | 13.28 | 14.95 | 13.885 | **12.97** | <u>13.227</u> |
| | HOMO (eV) | <u>0.316</u> | **0.311** | 0.329 | 0.349 | 0.318 | 0.337 |
| | LUMO (eV) | 0.443 | <u>0.394</u> | 0.409 | 0.443 | **0.389** | 0.411 |
| | Relative error | 1 | 0.98 | 1.1 | 1.04 | 0.96 | 1 |

*3.2 Similarity query results*

Based on the above reliable QSPR models, a similarity query method has been developed to design new molecules with similar characteristics to current promising molecules, such as FEC used as electrolyte additives for Li metal anodes. The capability of the similarity query method is ensured by a comprehensive database containing 280,000 molecules, carefully curated through multiple screening stages. The screening parameters include thermodynamic stability filtering (formation energy < 0 eV atom$^{-1}$) and synthetic accessibility evaluation (RAScore > 0.9 [35]).

Taking ethyl methyl carbonate (EMC), a widely used commercial electrolyte solvent, as an example, several new solvent molecules both linear carbonates and branched variants can be screening out (Figure 6). The new molecules share similar characteristics to EMC, i.e. $E_b$ = −1.53 eV, $\varepsilon$ = 1.34, $\eta$ = 5.5 mPa s, HOMO= −7.35 eV, and LUMO= 0.02 eV. Specifically, despite structural variations, all retrieved molecules maintain similar binding energies (from −1.29 to −1.68 eV) and HOMO levels (from −7.12 to −7.61 eV), suggesting these properties are primarily determined by the carbonate functional group. More importantly, several retrieved molecules are commercially available electrolytes, validating the practical utility of the search algorithm.

Query:

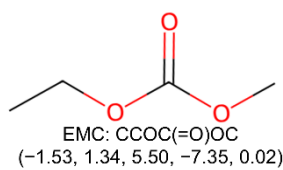

EMC: CCOC(=O)OC
(−1.53, 1.34, 5.50, −7.35, 0.02)

---

Results:

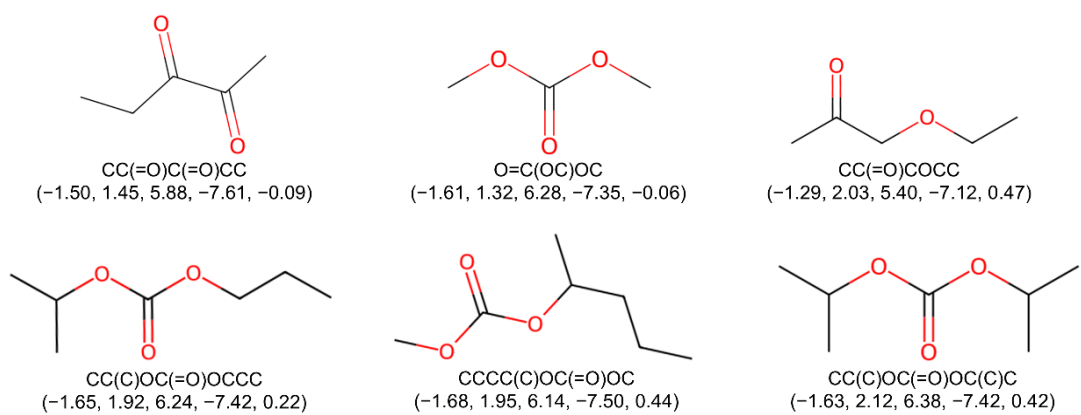

**Figure 6.** The query results of the EMC molecule. For each molecule, a 2D topology and a SMILES string are provided. The number of bottom the molecules is property vector ordered as Binding energy (eV), Dielectric Constant, Viscosity (mPa*s), HOMO (eV), and LUMO (eV).

*3.3 Property-targeted generation results*

Beyond the above query method, an AI generation model was further developed to design property-targeted molecules out of the embedded electrolyte database. Three distinct molecular generation tasks were specially investigated to demonstrate their practicability. (1) The generation of electrolyte molecules targeting a specific HOMO–LUMO gap; (2) The generation of molecules characterized by low binding energy and specific chemical formulas; (3) The generation of molecules exhibiting structural similarity to user-provided templates, represented as fingerprints.

Given the inherent stochasticity and potential instability associated with AI-generated molecules, an automated molecular cleaning workflow was established. The workflow serves as a critical post-generation filter, ensuring the quality and reliability of the generated molecular candidates. The workflow sequentially removes (1) molecules that fail topological checks, indicating structural inconsistencies; (2)

duplicate molecules, eliminating redundancy; (3) molecules already present in the training and validation sets, ensuring novelty; and (4) molecules with low synthetic accessibility scores (RAScore < 0.8), prioritizing molecules amenable to practical synthesis.

The evaluation of the generated molecules employs a suite of metrics tailored to the specific task. For property-targeted tasks (Tasks 1 and 2), state-of-the-art (SOTA) pre-trained models are utilized to predict the targeted properties, and the distribution of these properties is compared to that of the training set. The comparison provides a quantitative measure of the model's ability to generate molecules with the desired characteristics. For the fingerprint-targeted task (Task 3), the Tanimoto coefficient [36] serves as the primary metric, quantifying the structural similarity between the fingerprints of the generated molecules and the target fingerprints. This coefficient provides a direct assessment of the model's success in replicating the desired structural features.

The evaluation of property-targeted generation commences with Task 1, focusing on the generation of molecules with a predefined HOMO–LUMO gap. Two distinct generative model architectures are investigated: diffusion models, specifically the Energy-Based Diffusion Model (EDM [37]), and autoregressive models, represented by the conditional Graph-Schnet (cG-Schnet [38]). Initial comparative analysis reveals that conditional diffusion models, in this instance EDM, demonstrate superior performance in generating molecules with targeted HOMO–LUMO gaps, as evidenced in Figure 7. To illustrate this, consider the example of Dimethoxyethane (DME), a molecule absent from the original dataset yet residing in a relatively sparse region of the HOMO–LUMO gap distribution (Figure 7a). When the model is tasked with generating molecules within this region, the resulting molecules exhibit a concentrated distribution around the targeted area and successfully include DME (Figure 7 b), highlighting the model's sensitivity to conditional guidance and its ability to explore and populate sparse regions of the chemical space. This out-of-domain performance, the ability of the model to generate molecules with properties beyond the immediate scope of the training data, is a crucial indicator of its generalizability and practical utility in molecular design.

For Tasks 2 and 3, which involve generating molecules with low binding energy and specific chemical formulas, and generating molecules with structural similarity to target fingerprints, respectively, the investigation is narrowed to autoregressive models. The decision is predicated on the current absence of mature and readily adaptable diffusion models for these specific tasks. Within the autoregressive framework, variants of cG-Schnet are explored, specifically focusing on the impact of encoder architecture. The original encoder is replaced with both PaiNN and LEFTNet architectures. The results indicate that cG-LEFTNet exhibits superior performance on both Tasks 2 and 3. The enhanced performance is likely attributable to the superior encoding capabilities of LEFTNet, enabling it to more effectively capture and represent the intricate relationships between molecular structure and properties.

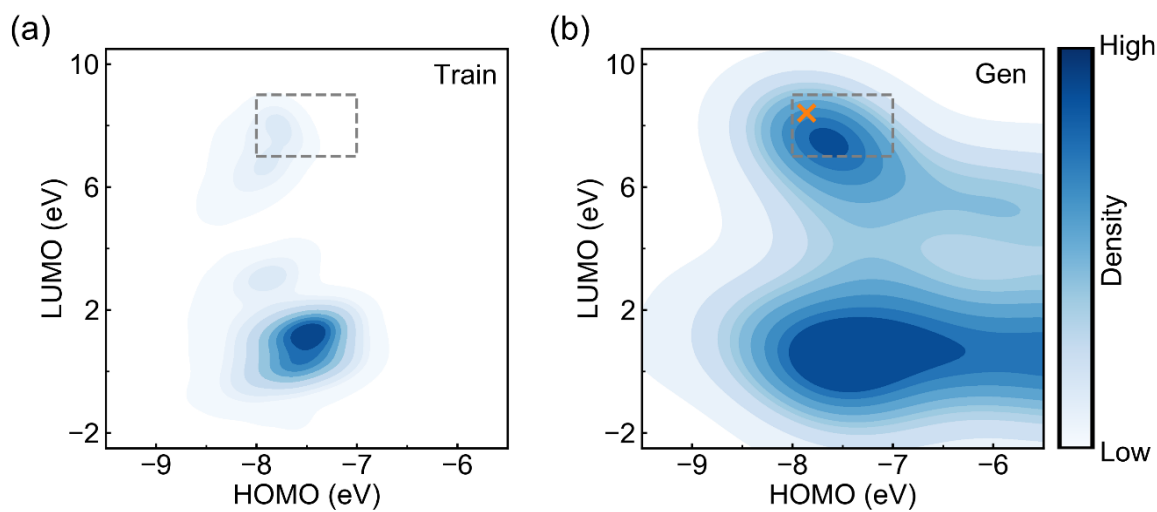

**Figure 7.** The out-of-domain performance of the generative model with respect to HOMO–LUMO properties. (a) The training dataset is sparse in the targeted area and does not contain the DMC molecule. (b) Whereas the generated dataset is relatively dense in the targeted area and has the DMC molecule.

*3.4 Retrosynthesis Validation*

Following the successful generation of molecules with targeted properties, the focus shifts to ensuring the practical feasibility of synthesizing these novel electrolyte candidates. A critical component of this feasibility assessment is the ability to predict viable synthetic pathways, a task typically addressed through retrosynthetic analysis. However, existing retrosynthesis methodologies are predominantly tailored for pharmaceutical molecules, relying heavily on template-based single-step predictors that often exhibit limited performance when applied to the structurally distinct domain of electrolytes. To overcome this limitation and enhance the applicability of retrosynthesis to electrolyte design, two key improvements are implemented.

First, the single-step predictor is replaced with the template-free model G2GT. The transition to a template-free approach is motivated by its potential for improved predictive capability outside the original training domain. To quantify the benefit of this substitution, a comparative analysis is performed against the established retrosynthesis method Askcos, utilizing an IID dataset comprising 100 electrolyte molecules. The performance of each method is evaluated based on its ability to correctly predict the immediate precursors (single-step prediction) for each target molecule, as measured by Top-k accuracy (where k = 1, 2, 3, 4, 5) and the number of molecules for which at least one correct precursor is identified within the top-k predictions ("Number of Molecules Recalled").

The above comparison results demonstrate that the G2GT predictor consistently outperforms Askcos in terms of Top-k accuracy across all values of k (Table 2). Specifically, G2GT achieves a Top-1 accuracy of 0.529 compared to Askcos's 0.452, indicating a higher probability of identifying the correct precursor in the first prediction. Furthermore, Askcos recalls a smaller number of molecules (8) compared to G2GT (17), suggesting G2GT can cover more electrolyte molecules. The combination of the G2GT single-step predictor and the Askcos path planner, termed G2GT-Askcos (with the workflow illustrated in Figure 4), leverages the strengths of both approaches, combining accurate single-step predictions with efficient path planning capabilities.

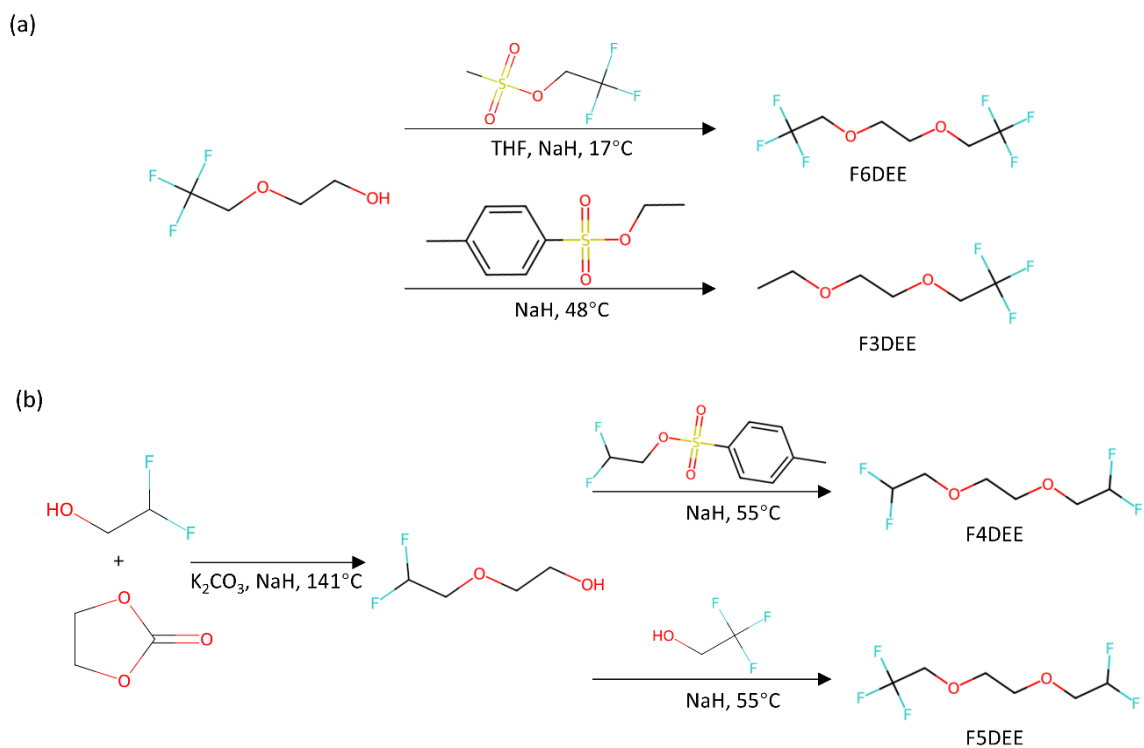

**Figure 8.** The retrosynthesis results of (a) F6DEE and F3DEE, (b) F4DEE and F5DEE are shown above. The proposed reagents are denoted above the reaction arrow, while the reaction conditions are labeled underneath.

**Table 2.** The comparison of retrosynthesis single-step prediction performance.

| Predictor | Top-1 Accuracy | Top-2 Accuracy | Top-3 Accuracy | Top-4 Accuracy | Top-5 Accuracy | Number of Molecules Recalled |
|---|---|---|---|---|---|---|
| G2GT | 0.529 | 0.636 | 0.682 | 0.708 | 0.725 | 17 |
| Askcos | 0.452 | 0.588 | 0.649 | 0.692 | 0.721 | 8 |

Second, to further enhance the performance of the retrosynthetic predictor within the electrolyte domain, the original USPTO reaction dataset is augmented. A subset of 1500 reactions is extracted from a larger pool of 1.1 million reactions in Reaxys, based on the structural similarity between the reaction products and common electrolyte molecules. This similarity is quantified using ECFP [39] fingerprints and the cosine similarity function. The selected reactions exhibit a minimum product similarity of 0.53 and an average similarity of 0.65, ensuring that the augmented dataset is relevant to the target chemical space.

Following fine-tuning with this augmented dataset, the G2GT single-step predictor demonstrates improved adaptation to the electrolyte domain. This adaptation is exemplified through a case study involving the retrosynthetic analysis of fluorinated diethoxyethane derivatives (F3DEE, F4DEE, F5DEE, and F6DEE). As a result, the reactions proposed by G2GT-Askcos for these molecules closely resemble those reported in the literature [40], validating the efficacy of the proposed retrosynthesis framework (Figure 8). In each case, the predicted reagents are indicated above the reaction arrow, and the corresponding reaction conditions are noted below. The successful prediction of plausible synthetic pathways for these representative electrolyte molecules underscores the potential of the G2GT-Askcos framework to facilitate the experimental validation and subsequent synthesis

*3.5 SEI Formation Analysis*

Following the validation of synthetic accessibility, the investigation turns to predicting the impact of the designed electrolyte molecules on battery performance, specifically focusing on the formation of the solid electrolyte interphase (SEI). Understanding SEI composition is crucial for designing electrolytes that promote stable battery cycling. To predict SEI formation pathways, a modified version of the HiPRGen software, termed xHiPRGen(extended HiPRGen), is employed. Key enhancements to HiPRGen include expanding the built-in dataset to encompass the entire LiBE electrolyte database [26] and automating the database construction process for user customization (Figure 5). These modifications allow for the exploration of SEI formation from a broader range of electrolyte chemistries, including those containing fluorine, nitrogen, phosphorus, and sulfur.

To demonstrate the utility of xHiPRGen for predicting SEI formation pathways, a case study is conducted focusing on the decomposition of fluoroethylene carbonate (FEC), a widely used electrolyte additive known to promote the formation of stable SEI layers. The predicted decomposition pathway of FEC, as generated by xHiPRGen, is presented in Figure 9. The Gibbs free energy change associated with each reaction step is indicated above the reaction arrow, providing a thermodynamic assessment of the reaction feasibility.

The analysis begins with the initial coordination of FEC with a lithium ion (Li-FEC). xHiPRGen predicts that the transition from a single coordination state to a double coordination state (Figure 9a) is thermodynamically favorable, with the double coordination complex exhibiting greater stability. This prediction aligns with previous computational and experimental studies [41] that have identified the double coordination state of Li-FEC as a key intermediate in SEI formation.

Following the formation of the stable double coordination complex, xHiPRGen predicts a ring-opening reaction of Li-FEC (Figure 9b), leading to the formation of an open-chain intermediate. This intermediate then interacts with another double-coordinated Li-FEC molecule, resulting in the formation of a more stable dimeric intermediate (Figure 9c). Finally, these dimeric intermediates undergo further reactions, including the release of lithium fluoride (LiF), leading to the formation of a polymeric SEI component (Figure 9d). The predicted polymeric structure is consistent with experimental observations of SEI layers formed from FEC-containing electrolytes, which often exhibit a significant fraction of polymeric and oligomeric species.

The case study vividly illustrates the potential of xHiPRGen to provide valuable insights into the complex decomposition pathways of electrolyte molecules and the resulting SEI composition. The ability to predict the potential SEI products formed under operational conditions offers a powerful tool for guiding the design and optimization of electrolyte formulations, enabling the development of batteries with enhanced performance and longevity. Furthermore, the automated database construction and expansion capabilities of xHiPRGen make it readily adaptable to the exploration of novel electrolyte chemistries and the investigation of SEI formation in emerging battery technologies.

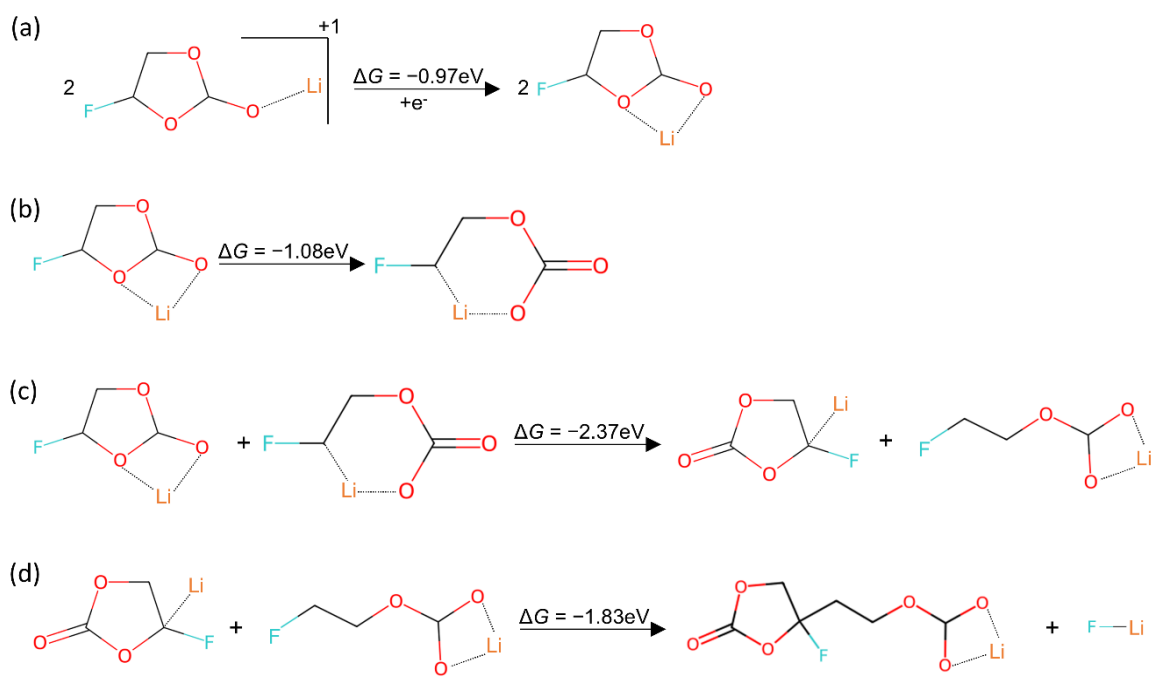

**Figure 9.** The proposed FEC decomposition route to potential SEI product. The Gibbs free energy of each reaction is denoted on top of the reaction arrow.

## 4. Conclusions

An AI platform, namely the Uni-Electrolyte, has been developed to design battery electrolyte molecules based on a data-driven approach. The Uni-Electrolyte integrates three synergistic module, i.e. EMolCurator, EMolForger, and EMolNetKnittor. EMolCurator, the AI-assisted molecular design framework, leverages QSPR models, multi-criteria screening, similarity searches, and AI-driven molecular generation to efficiently explore the vast chemical space and identify promising electrolyte candidates with tailored properties. EMolForger, the AI-powered retrosynthetic analysis module, bridges the gap between computational design and experimental synthesis by providing optimized synthetic routes and detailed reagent information. Last but not least, EMolNetKnittor, the comprehensive SEI formation analysis platform, enables detailed investigation of the complex interfacial processes that govern battery performance, providing crucial insights into the SEI composition and formation mechanisms. The integration of the three modules within the Uni-Electrolyte platform affords a powerful and versatile toolset for both electrolyte researchers and engineers, enabling an AI-driven design of advanced electrolytes for next-generation batteries.

**Acknowledgments**

This work was supported by the National Natural Science Foundation of China (T2322015, 92472101, 22393903, 22393900, and 52394170), the National Key Research and Development Program (2021YFB2500300), and the Beijing Municipal Natural Science Foundation (L247015 and L233004).